\documentclass[prb,twocolumn,superscriptaddress,longbibliography,floatfix]{revtex4-2}
\usepackage{graphicx,amsfonts,amssymb,amsmath,xspace,dsfont}
\usepackage[colorlinks=true,citecolor=BurntOrange,linkcolor=red,urlcolor=Bittersweet]{hyperref}
\usepackage{color}
\usepackage[dvipsnames]{xcolor}
\definecolor{g}{rgb}{.1,0.4,.1} 
\definecolor{b}{rgb}{0,0.2,1}
\definecolor{rouge}{rgb}{0.82,0.,0.}
\definecolor{vert}{rgb}{0.,0.82,0.}
\definecolor{orange}{rgb}{1,0.5,0.}
\definecolor{bleu}{rgb}{0.,0.,0.82}
\definecolor{m}{rgb}{0.82,0.,0.82}
\definecolor{vert2}{rgb}{0.,0.5,0.}
\definecolor{rougeclair}{rgb}{1.0,0.7,0.7}

\newcommand{\be}{\begin{equation}}
\newcommand{\ee}{\end{equation}}
\newcommand{\beqn}{\begin{eqnarray}}
\newcommand{\eeqn}{\end{eqnarray}}

\begin{document}

\title{Kitaev model in regular hyperbolic tilings}

\author{Julien Vidal}
\email{julien.vidal@sorbonne-universite.fr}
\affiliation{Sorbonne Universit\'e, CNRS, Laboratoire de Physique Th\'eorique de la Mati\`ere Condens\'ee, LPTMC, F-75005 Paris, France}

\author{R\'emy Mosseri}
\email{remy.mosseri@sorbonne-universite.fr}
\affiliation{Sorbonne Universit\'e, CNRS, Laboratoire de Physique Th\'eorique de la Mati\`ere Condens\'ee, LPTMC, F-75005 Paris, France}


\begin{abstract}
We study the Kitaev model on regular hyperbolic trivalent tilings. Depending on the length $p$ of the elementary polygons, we examine two distinct tri-colorings of the tiling. Using a recent conjecture on the ground-state flux sector, we compute the phase diagram via exact diagonalizations and derive analytical expressions for the effective Hamiltonians in the isolated-dimer limit which are valid for all values of $p$. Our results interpolate between the Euclidean honeycomb lattice and the trivalent Bethe lattice ($p=\infty$) for which we derive the exact solution of the phase boundaries.   
\end{abstract}

\maketitle

%
%
\section{Introduction}
\label{sec:intro}
%
%
Following the discovery of the fractional quantum Hall effect~\cite{Tsui82,Stormer99}, much attention has been paid to two-dimensional systems hosting anyons~\cite{Leinaas77,Wilczek82_1,Wilczek82_2,Goldin85}. These collective excitations are known to obey nontrivial braiding statistics which have been probed experimentally in recent years~\cite{Bartolomei20,Nakamura20}. Classifying and understanding anyonic (topological) quantum phases of matter is one of the most challenging problems in condensed matter physics~\cite{Wen17}, notably because of their potential use in topological quantum computation~\cite{Nayak08}.  

An important step was achieved in this direction by Kitaev~\cite{Kitaev06} who proposed an exactly solvable spin model and the celebrated sixteenfold way which describes all possible phases of Majorana fermions coupled to a $\mathbb{Z}_2$ gauge field. A remarkable feature of this model is its ability to generate both  Abelian  (toric code) and non-Abelian (Ising) anyons when varying the Hamiltonian parameters. 
Many variations of this model have been proposed by changing, e.g.,  the geometry, the space dimension, the Hilbert space dimension, the interactions (see, e.g., Refs.~\cite{Trebst22,Matsuda25} for recent reviews). However, to ensure the exact solvability of this model based on a free Majorana-fermion representation of the Hamiltonian, one important constraint is to consider trivalent structures.  

Recently, a new extension of the Kitaev model has emerged by considering tilings of the hyperbolic plane, i.e., a space with constant negative curvature. Latest works on trivalent tilings made of  $7$-gons~\cite{Mosseri25}, $8$-gons~\cite{Lenggenhager25}, and $9$-gons~\cite{Dusel25} motivated us to explore the Kitaev model on any regular hyperbolic tilings $\{p,3\}$, i.e., on regular tesselations of the hyperbolic plane by a trivalent tiling made of $p$-gons. The honeycomb lattice $(p=6)$ corresponds to the limiting Euclidean (zero curvature) case~\cite{Kitaev06} whereas tilings with $p<6$ (not considered here) correspond to trivalent Platonic solids [tetrahedron $(p=3)$, cube $(p=4)$, and dodecahedron $(p=5)$]. Our main goal is to connect these recent results\cite{Mosseri25,Lenggenhager25,Dusel25}, with  the $p=\infty$ limit, which corresponds to the trivalent Bethe lattice~\cite{Mosseri82}. As we will see, this limit can be reached in various ways depending on the coloring considered. Using numerical exact diagonalizations (ED) of the Hamiltonian, we compute the phase diagram of the Kitaev model for finite values of $p$ (up to 16). We also derive the effective  low-energy Hamiltonian for arbitrary $p$ in the isolated-dimer limit, in which the system is gapped and in a toric code phase. Finally, we present the exact phase diagram for the Bethe lattice by revisiting (and correcting) the results given in Ref.~\cite{Kimchi14}. 

This paper is organized as follows: In Sec.~\ref{sec:model}, we introduce the Kitaev model and the two colorings (coupling patterns) considered in this study, namely, the Kekul\'e coloring and the Kitaev coloring. The former allows us to study $\{p,3\}$ tilings with $p\mod 2=0$ and the latter is only relevant for when $p \mod 3=0$.  Section~\ref{sec:gss} discusses the conjecture proposed in Refs.~\cite{Lieb93,Cassella23} that determines the ground-state sector, and Sec.~\ref{sec:ABfluxes} provides some remarks about the gauge choice and Aharonov-Bohm (AB) fluxes. Section~\ref{sec:diagrams} is dedicated to the numerical study of the fermion gap for the  aforementioned families of tilings. In Sec.~\ref{sec:dimer}, we use the formalism given in Ref.~\cite{Petrova14} to derive the low-energy effective Hamiltonians in the isolated-dimer limit allowing for a direct calculation of the vison gap in this limit.  
The $p=\infty$ case, which corresponds to the trivalent Bethe lattice is discussed in Sec.~\ref{sec:Bethe} where we compute the exact phase diagram within the full parameter range. We conclude and provide some perspectives in Sec.~\ref{sec:conclusion}. Technical details are given in Appendix~\ref{app:Bethe}. 
In Appendix~\ref{app:Hecke}, we present an alternative approach towards the Bethe lattice limit by examining the Kitaev model on certain polygonal tree structures. These tilings are related to the Hecke group Cayley graphs and allow for an exact solution that matches that of the Bethe lattice one in the case of infinite-length polygons.

%
%
\section{Model}
\label{sec:model}
%
%

For any three-edge colorable tiling embedded on an orientable surface, the Kitaev model Hamiltonian is defined as 
%
%
\begin{equation}
  \label{eq:ham0}
  H=-\sum_{(j,k)}
  J_{\alpha} \, \sigma_j^\alpha\sigma_k^\alpha,
\end{equation} 
%
%
where the $\sigma_i^\alpha$'s are the Pauli matrices acting on the spin located at site $i$. In Eq.~\eqref{eq:ham0}, the sum runs over all links $(j,k)$ of the tiling, and the type $\alpha=x,y,$ or $z$ is given by the corresponding edge color. For any plaquette $n$, the operator 
%
%
\begin{equation}
W_n= \prod_{(j,k)\in {\partial n}} \sigma_j^\alpha\sigma_k^\alpha,
\label{eq:wpdef}
\end{equation} 
%
%
is a conserved quantity, i.e., $[H,W_n]=0$~\cite{Kitaev06}.  Here, the product runs over all links $(j,k)$ of the oriented \mbox{plaquette} $n$ (e.g., clockwise). Additionally, one has $[W_n,W_{n'}]=0$, for any plaquettes $n$ and $n'$, which allows one to block diagonalize the Hamiltonian according to each map of the $W_n$'s. 
%
%
\begin{figure}[t]
\centering
\includegraphics[width=.7\columnwidth]{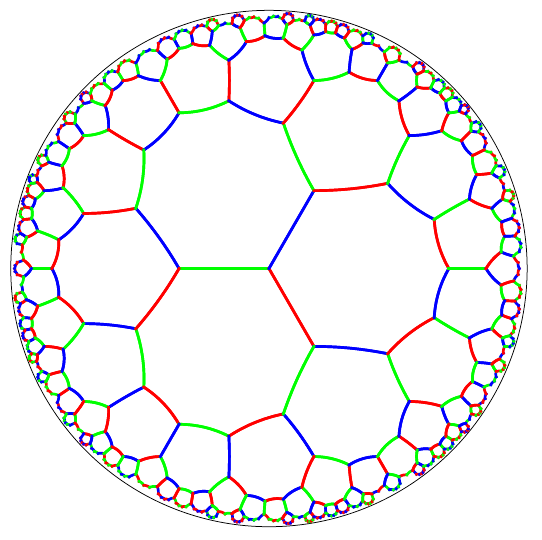}
\caption{A piece of the $\{8,3\}$ tiling in the Poincar\'e disk conformal representation with Kekul\'e coloring, in which each polygon is only made up of two colors. The red, green, and blue links are associated with $J_x$, $J_y$, and $J_z$ couplings, respectively.}
\label{fig:disk_8}
\end{figure}
%
%
For a given configuration of the $W_n$'s, the Hamiltonian \eqref{eq:ham0} can be mapped onto a quadratic Majorana fermion Hamiltonian
%
%
\be 
H=\frac{\mathrm{i}}{4} \sum_{j,k} A_{jk} c_j c_k,
\label{eq:ham_Majo}
\ee  
%
%
where the sum runs over all sites of the tiling, and where  $c_j$ is a Majorana operator acting on site $j$. As explained by Kitaev, this mapping leads to an extended Hilbert space and a projection onto the physical subspace must be performed~\cite{Kitaev06}.  
The matrix $A$ is a real skew-symmetric matrix and its elements $A_{jk}=2 \, J_{\alpha} \, u_{jk}$ depend on $\mathbb{Z}_2$ variables \mbox{$u_{jk}=-u_{kj}=\pm 1$}, which are defined on each link of the lattice~\cite{Kitaev06}. In this language, the eigenvalue of the operator $W_n$ is given by
%
%
\be
w_n=(-{\rm i})^p \prod_{(j,k)\in {\partial n}} u_{jk},
\label{eq:fluxdef}
\ee
%
%
where $p$ is the number of edges of the plaquette $n$, and the product again runs over all links $(j,k)$ of the oriented plaquette $n$~\cite{Kitaev06,Petrova14}.
Thus, for each configuration of the gauge variables $u_{jk}$'s, the problem reduces to a quadratic Majorana fermion Hamiltonian. Note that on a closed surface, one always has $\prod_n W_n=\mathds{1}$, or equivalently $\prod_n w_n=1$.
Importantly, the Hamiltonian \eqref{eq:ham0} is time-reversal symmetric. However, if the tiling contains plaquettes of odd length, time-reversal symmetry (TRS) is spontaneously broken~\cite{Kitaev06} and the ground state is doubly degenerate (Kramers pairs)~\cite{Kramers31}. This phenomenon was first analyzed in the decorated honeycomb lattice~\cite{Yao07,Dusuel08_2}.

\begin{figure}[t]
\centering
\includegraphics[width=.7\columnwidth]{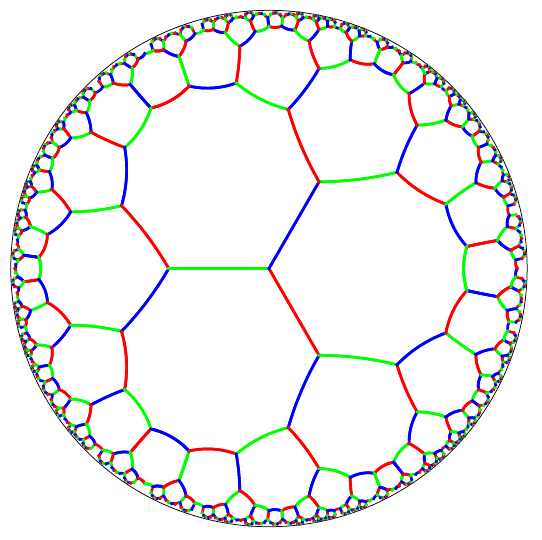}
\caption{A piece of the $\{9,3\}$ tiling in the Poincar\'e disk conformal representation with Kitaev coloring, in which each polygon is a periodic arrangement of the three colors red, green, and blue). The color code is the same as in Fig.~\ref{fig:disk_8}.}
\label{fig:disk_9}
\end{figure}

\begin{figure}[t]
\centering
\includegraphics[width=.7\columnwidth]{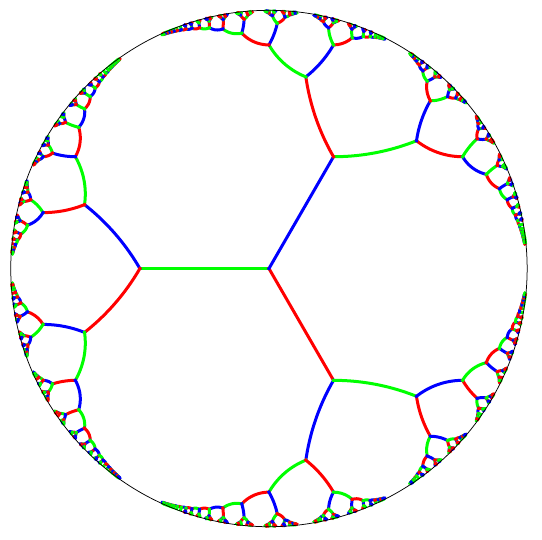}
\caption{A piece of the $\{\infty,3\}$ tiling (trivalent Bethe lattice) in the Poincar\'e disk conformal representation, for which the three-edge coloring is irrelevant. The color code is the same as in Fig.~\ref{fig:disk_8}.}
\label{fig:disk_Bethe}
\end{figure}

The properties of the Kitaev model for a given tiling strongly depend on the coloring considered (see Ref.~\cite{Kamfor10} for a discussion in the honeycomb lattice). In particular, when $p \mod 2=0$, one can use a ``Kekul\'e coloring" (see Fig.~\ref{fig:disk_8}) where each $p$-gon has only two types of links (see Refs.~\cite{Kamfor10} for $p=6$,  and Ref.~\cite{Lenggenhager25} for $p=8$). When $p \mod 3 =0$, one can use the ``Kitaev coloring"~\cite{Kitaev06} (see Fig.~\ref{fig:disk_9}), which consists of  alternating $x$, $y$, and $z$ couplings around each $p$-gon. This coloring was recently analyzed in the $\{9,3\}$ tiling~\cite{Dusel25}.  
In the following, we will focus on these two families of tilings, studying how, when $p$ goes to infinity, the corresponding phase diagrams converge to that of the trivalent Bethe lattice (see Fig.~\ref{fig:disk_Bethe}) for which the coloring is irrelevant. Note that other values of $p$ may be of interest, e.g., \mbox{$p=7$~\cite{Mosseri25}} but they do not allow for either of the two colorings mentioned above. Additionally, when periodic boundary conditions are considered (closed surface), frustration effects, because of odd length systoles, may occur preventing the aforementioned coloring from being realized (see Ref.~\cite{Lenggenhager25} for further discussions). 

%
%
%
%

%
%
\section{Ground-state sector(s)}
\label{sec:gss}
%
%

Among all possible gauge choices, the ground-state manifold has  recently been conjectured to be obtained when
%
%
\be
w_n^{{\rm g.s.}}=-(\pm {\rm i})^p,
\label{eq:fluxmin}
\ee
%
%
for each elementary plaquette $n$ with $p$ edges, provided the system is not too small~\cite{Cassella23}. First proposed by Lieb and Loss~\cite{Lieb93}, this conjecture is in agreement with recent works on disordered tilings systems (see also Ref.~\cite{Grushin23}),  and is supposed to hold for any coloring and nonnegative couplings $J_\alpha$. As an aside, let us stress that when one of the couplings dominates (e.g., when $J_z \gg J_x,J_y$),  the perturbative results derived in Ref.~\cite{Petrova14} corroborate this conjecture.

For even $p$, the ground-state configuration is thus given by $w_n^{{\rm g.s.}}=+1$ when $p \mod 4=2$, and by $w_n^{{\rm g.s.}}=-1$ when $p \mod 4=0$. For odd $p$, this conjecture predicts two ground-state sectors (Kramers pair~\cite{Kramers31}) associated with $w_n^{{\rm g.s.}}=\pm {\rm i}$) reminiscent of the spontaneous TRS breaking expected in this case (see the above discussion).

In the trivalent Bethe lattice (infinite-$p$ limit) where loops are absent, it is important to underline that all gauge configurations and all colorings lead to the same spectrum. This issue was originally addressed in Ref.~\cite{Kimchi14}, but, as we will see below, some solutions were overlooked.

%
%
%
%

%
%
\section{Closed surfaces and Aharonov-Bohm fluxes} 
\label{sec:ABfluxes}
%
%
To determine the phase diagram of the Kitaev model on a given $\{p,3\}$ tiling, we performed exact diagonalizations (ED) of the Hamiltonian \eqref{eq:ham_Majo} in the ground-state sector(s) given by the condition \eqref{eq:fluxmin}. To limit finite-size effects, for a fixed $p$, we considered the cluster with the largest number of exact moments of its adjacency matrix (as compared to the infinite tiling)~\cite{Mosseri23} in the list of trivalent graphs with up to $N_{\rm v}=10 000$ vertices (see Ref.~\cite{Conder06}). Each cluster corresponds to a closed surface (with periodic boundary conditions) of genus $g=1+\frac{p-6}{4p}N_{\rm v}$ so that no boundary effects are present. 
Consequently, the results depend not only on the flux $\phi_n$ in each plaquette ($w_n={\rm e}^{{\rm i} \phi_n}$) but also on the AB fluxes in each noncontractible loop. Once these quantities are fixed, the spectrum of the original spin Hamiltonian is ultimately found by selecting the physical states from the extended Majorana fermion Hamiltonian~\cite{Pedrocchi11}. For sufficiently large system sizes, the spectrum becomes dense and this projection step is not important to determine the actual value of the gap. For the clusters studied here, the estimated error is smaller than a few percents. Moreover, this spectrum is also sensitive to finite-size effects, which must be analyzed carefully (see Refs.~\cite{Stegmaier22,Mosseri25} for related discussions). 

For a given tiling on a closed surface, the exact fermion gap $\Delta_{\rm f}$ defined as the smallest nonnegative (physical) eigenvalue of the matrix ${\rm i}A$, is obtained by minimizing over all possible AB fluxes. In practice, computing the fermion gap is impossible because of the large number AB fluxes to be sampled ($g$  grows as $N_{\rm v}$). Therefore, for a given gauge choice [fixing both local ($w_n$) and nonlocal (AB) fluxes], one can only obtain an upper bound for the fermion gap.

%
%
\section{Phase diagrams} 
\label{sec:diagrams}
%
%
\subsection{Kekul\'e-colored tilings $(p\mod 2=0)$}

\begin{figure}[t]
\includegraphics[width=\columnwidth]{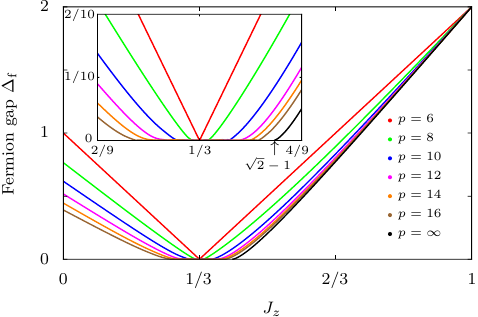}
\caption{The fermion gap $\Delta_{\rm f}$ for the Kekul\'e coloring is shown as a function of $J_z$ along the line $J_x=J_y$,  with the constraint \mbox{$J_x+J_y+J_z=1$}, which fixes the energy unit. The inset shows a zoom around $J_z=1/3$. The results were obtained from ED of {\it ad hoc} clusters~\cite{Conder06} with a gauge configuration realizing the ground-state flux sector [see Eq.~\eqref{eq:fluxmin}], except for $p=\infty$ where the exact solution discussed in Sec.~\ref{sec:Bethe} was used. When $J_z=1$ ($J_x=J_y=0$), the isolated-dimer limit is recovered and $\Delta_{\rm f}=2$  (see Sec.~\ref{sec:dimer}). For $J_z=0$, the isolated $p$-gon limit is obtained, and the gap is given by Eq.~\eqref{eq:gap}. In the Bethe lattice ($p=\infty$), one has $\Delta_{\rm f}=0$ for $J_z\in [0,\sqrt{2}-1$] (see Sec.~\ref{sec:Bethe}).
}
\label{fig:gap_Ke}
\end{figure}

As explained in Sec.~\ref{sec:model}, when $p\mod 2=0$, one can use the Kekul\'e coloring to three-color the corresponding $\{p,3\}$ tiling. For $p=6$, the phase diagram was exactly computed in Ref.~\cite{Kamfor10} consisting of a single gapless point found for $J_x=J_y=J_z$. For any other values of the $J_\alpha$'s, the system is gapped. The phase diagram for $p=8$ has been analyzed in Ref.~\cite{Lenggenhager25} and the authors conclude that the phase diagram may be the same as for $p=6$ using various approaches. Given that when $p$ goes to infinity, one should recover the diagram of the trivalent Bethe lattice (see Sec.~\ref{sec:Bethe}), this result seems somewhat surprising. Therefore, we revisited this issue by means of ED and we observed a small gapless phase near the isotropic point. This is most evident plotting the gap along the line parametrized by  $J_x+J_y+J_z=1$ and $J_x=J_y$ (see the dashed line in Fig.~\ref{fig:gap_Ke}) where the gap clearly vanishes in a finite region. As mentioned earlier, the boundaries of this gapless phase are a bit imprecise (about a few percent) because we cannot sample all possible AB fluxes. However, for any fixed gauge choice, we always observe a finite gapless phase. As $p$ increases, the gapless phase grows and converges to that of the Bethe lattice (see Fig.~\ref{fig:gap_Ke} and Sec.~\ref{sec:Bethe}).

When one of the couplings vanishes (e.g., $J_z=0$), the system consists of a set of isolated $p$-gons made of alternating couplings ($J_x,J_y$). When $J_x=J_y=1/2$, the fermion gap is then given by
%
%
\be
\Delta_{\rm f}=\min\{|2\cos(2\pi j/p+\phi)|,j=0,\dots,p-1\}.
 \label{eq:gap}
 \ee
%
%
The phase shift ($\phi=\pi/p$, if $p \mod 4=0$, and $0$ otherwise) comes from the flux constraint [see Eq.~\eqref{eq:fluxmin}]. As can be seen in Fig.~\ref{fig:gap_Ke}, this gap is always finite but as $p$ approaches infinity $\Delta_{\rm f}=0$, which is  in agreement with the Bethe lattice results discussed in Sec.~\ref{sec:Bethe}.

%
%
%
%

\subsection{Kitaev-colored tilings $(p\mod 3=0)$}

\begin{figure}[t]
\includegraphics[width=\columnwidth]{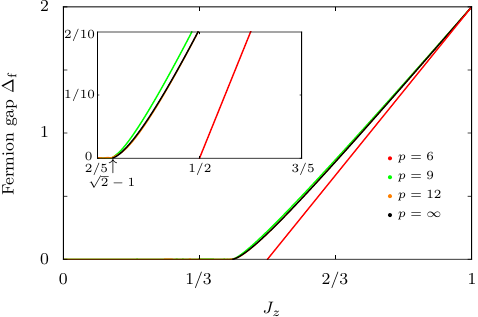}
\caption{The fermion gap $\Delta_{\rm f}$ for the Kitaev coloring is shown as a function of $J_z$ along the line $J_x=J_y$,  with the constraint \mbox{$J_x+J_y+J_z=1$} (see also the caption for Fig.~\ref{fig:gap_Ke}). Inset: A zoom around $J_z=1/3$. Unlike the Kekul\'e coloring, the phase diagram with the Kitaev coloring quickly converges to that  of the Bethe lattice as $p$ increases. }
\label{fig:gap_Ki}
\end{figure}

Another way to reach the Bethe lattice limiting case is to consider the set of $\{p,3\}$ tilings where $p\mod 3=0$, which allow for a Kitaev coloring. For $p=6$, the system is gapless when $J_\alpha \leqslant J_\beta+J_\gamma$, giving rise to the triangular-shaped phase originally computed by Kitaev~\cite{Kitaev06}. For $p=9$, the phase diagram was recently obtained numerically in Ref.~\cite{Dusel25} and the extension of the gapless phase is found to be slightly smaller than for $p=6$. Again, this is consistent with the idea that, at large $p$, one should converge to the Bethe lattice phase diagram discussed in Sec.~\ref{sec:Bethe}. This convergence is clearly observed in Fig~\ref{fig:gap_Ki} where we plot the gap for $p=6, 9,12$, and $\infty$, along the cut shown in Fig.~\ref{fig:diagram} (dashed line).
%
%
%
%

When one of the couplings (e.g., $J_z$)  vanishes, the system consists of a set of decoupled chains with alternating couplings ($J_x$, $J_y$). The fermion gap $\Delta_{\rm f}$ is always finite, but it vanishes when $J_x=J_y$.

%
%
\section{The isolated-dimer limit} 
\label{sec:dimer}
%
%
To analyze the nature of the gapped phase, we adopt the original approach proposed by Kitaev and examine the isolated-dimer limit in which one of the coupling dominates. In this limit, one can derive an effective low-energy Hamiltonian that describes weakly coupled dimers. Using the results from Petrova {\it et al.}~\cite{Petrova14} (valid for any graph) for Kekul\'e-colored tilings (\mbox{$p \mod 2=0$}) and assuming \mbox{$J_z\gg J_x, J_y$}, one gets at leading order $p/2$, 
%
%
\beqn
H_{\rm eff}&=&{\rm const}+\frac{(-1)^{p/2}\Gamma\left(\tfrac{p-1}{2}\right)}{\sqrt{\pi} \: \Gamma\left(\tfrac{p}{2}\right)} \frac{1}{J_z^{p/2-1}} \times  \nonumber\\
&&{\Bigg(}J_x^{p/2}\sum_{n,(xz)} W_{n}+J_y^{p/2}\sum_{n,(yz)} W_{n} {\Bigg)},
\label{eq:hameff1}
\eeqn
%
%
where the first (second) sum runs over all plaquettes $n$ made of $x$ and $z$ ($y$ and $z$) links (see Fig.~\ref{fig:disk_8}). The standard  Euler-Gamma function is denoted by $\Gamma$. It is easy to check that expression \eqref{eq:hameff1} reproduces the results obtained in Ref.~\cite{Kamfor10} for $p=6$, and in Ref.~\cite{Lenggenhager25} for $p=8$. 

For tilings with a Kitaev coloring (\mbox{$p \mod 3=0$}), one must distinguish between two cases depending on the parity of $p$. When $p$ is even ($p\mod 6=0$), the effective Hamiltonian at leading order $2p/3$ is given by
%
%
\beqn
H_{\rm eff}&=&{\rm const}+\frac{(-1)^{p/2} \Gamma\left(\tfrac{p-1}{2}\right) \Gamma\big(\tfrac{1+p/3}{2}\big)}{\pi \: \Gamma\left(\tfrac{2p}{3}\right)}  \frac{J_x^{p/3} J_y^{p/3}}{J_z^{2p/3-1}}\times \nonumber\\
&& {\Bigg(}  \sum_{n} W_{n} {\Bigg)},
\label{eq:hameff2}
\eeqn
%
%
where the sum runs over all plaquettes $n$. For $p=6$, this expression matches the one found by Kitaev~\cite{Kitaev06} (see also Ref.~\cite{Vidal08_2} for higher-order terms). By contrast, when $p$ is odd ($p\mod 6=3$), one gets the following form at leading order $4p/3-2$:
%
%
\beqn
H_{\rm eff}&=&{\rm const}+\frac{4 \: \Gamma\left(p-\tfrac{3}{2}\right) \Gamma\big(\tfrac{p}{3}+\tfrac{1}{2}\big)}{ \pi \: \Gamma\left(\tfrac{4p}{3}-1\right)}  \frac{J_x^{2p/3-2} J_y^{2p/3-2}}{J_z^{4p/3-3}}\times \nonumber\\
&&   {\Bigg(} J_x^2 \sum_{(n,n')_y} W_{n} W_{n'}+J_y^2 \sum_{(n,n')_x} W_{n} W_{n'} {\Bigg)},
\label{eq:hameff3}
\eeqn
%
%
where the first (second) sum runs over all pairs plaquettes $(n,n')$ sharing a $y$-link  ($x$-link) (see Fig.~\ref{fig:disk_9}). For \mbox{$p=9$}, this result coincides with the one given in Ref.~\cite{Dusel25}, as expected. Notably, we observe that the leading-order terms involve two neighboring plaquettes. This is reminiscent from the spontaneous TRS breaking that occurs for odd $p$,  where the two ground-state sectors correspond to uniform flux maps ($w_n=\pm {\rm i}$, for all $n$).

These effective Hamiltonians corroborate the conjecture given in Eq.~\eqref{eq:fluxmin}. They also allow to straightforwardly compute the vison gap, $\Delta_{\rm v}$, obtained by exciting plaquettes (see Refs.~\cite{Mosseri25,Dusel25} for recent discussions of this gap).  Note that for $J_x=J_y=0$, the fermion gap, denoted by $\Delta_{\rm f}$, is obtained by exciting a $z$-dimer and is equal to $2 J_z$.
The gapped phase adiabatically connected to the isolated-dimer limit has a Chern number $\nu=0$ (see Ref.~\cite{Kitaev06}). According to the 16-fold-way classification, this corresponds to a toric code phase~\cite{Kitaev06}.

%
%
\section{The trivalent Bethe lattice} 
\label{sec:Bethe}
%
%
As the size $p$ of the elementary plaquettes approaches infinity, the tiling becomes equivalent to the trivalent Bethe lattice. Each site is threefold coordinated and there are no closed loops of finite length in the system. In this limit, the spectrum of the Kitaev model is independent of the coloring and independent of the $u_{jk}$'s. In fact,  the notion of $w_n$'s becomes irrelevant. This problem was originally studied in Ref.~\cite{Kimchi14} as a large-loop limit of a three-dimensional hyperhoneycomb lattice (see also Ref.~\cite{Mandal09}). 

To determine the phase boundaries of the gapless phase, we follow Ref.~\cite{Kimchi14} (see Appendix D) and consider the following set of equations derived in Appendix~\ref{app:Bethe}:
%
%
\be
1= \sum_{\alpha=x,y,z} \varepsilon_\alpha \sqrt{1-J_\alpha^2 r^2},
\label{eq:green3}
\ee
%
%
 where $r=2\pi \rho (0)$ [$\rho(E)$ being the normalized density of states of the matrix ${\rm i} A/2$ at energy $E$], and  $\varepsilon_\alpha= \pm 1$. Different solutions of Eq.~\eqref{eq:green3} are parametrized by $\varepsilon_\alpha$'s. In Appendix D of Ref.~\cite{Kimchi14}, the authors only consider the case $\varepsilon_\alpha= +1$ for all $\alpha$ and overlooked the other cases which are relevant to determine the exact phase boundaries.  
Indeed, let us seek for a solution of Eq.~\eqref{eq:green3} with infinitesimal $r$. We can then expand Eq.~\eqref{eq:green3} at small $r$ and obtain
%
%
\be
1= \sum_{\alpha=x,y,z} \varepsilon_\alpha \left(1-\frac{J_\alpha^2 r^2}{2}\right)+O(r^4).
\label{eq:green4}
\ee
%
%
Determining the region where the gap vanishes is equivalent to finding the set of parameters $J_\alpha$ for which $r=0$. As discussed in Appendix D of Ref.~\cite{Kimchi14}, this never happens with the choice \mbox{$\varepsilon_\alpha=+ 1$ for all $\alpha$},  but this is possible with other sign choices as we shall now discuss.

%
%
\begin{figure}[t]
\includegraphics[width=0.7\columnwidth]{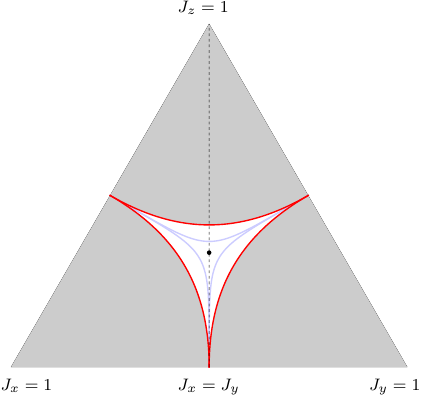}
\caption{Phase diagram of the Kitaev model on the trivalent Bethe lattice ($J_x+J_y+J_z=1$). Gapped phases are displayed in gray. The exact boundaries of the gapless phase (red lines) are given by: $J_\alpha^2=J_\beta^2+J_\gamma^2$. The blue lines are the boundaries obtained with $\varepsilon_\alpha=+1$~\cite{Kimchi14} [see Eq.~\eqref{eq:Kimchi2}]. The isotropic point \mbox{$J_x=J_y=J_z$}, is highlighted with a black dot. The dashed line corresponds to the cut along which the gap is computed in Figs.~\ref{fig:gap_Ke} and \ref{fig:gap_Ki}.
}
\label{fig:diagram}
\end{figure}
%
%
Now, let us assume that we selected, in Eq.~\eqref{eq:green4}, the solution corresponding to, e.g., \mbox{$\varepsilon_x=\varepsilon_y=-\varepsilon_z=+1$}. Then, it is straightforward to see  that Eq.~\eqref{eq:green4} is satisfied when
%
%
\be 
J_z^2= J_x^2+J_y^2.
\ee 
%
%
This equation and its counterparts, which are obtained by permuting the indices, define the exact critical lines that separate the gapless phase ($r \neq 0$) from the gapped phase ($r= 0$) (see Fig.~\ref{fig:diagram}) for the Kitaev model on the trivalent Bethe lattice. Surprisingly, the corresponding phase diagram shown in Fig.~\ref{fig:diagram} is exactly the same as the one the Kitaev model on the honeycomb lattice in the vortex-full sector~\cite{Pachos07}!

For completeness, we also  determine the boundaries of the gapless phase obtained by considering only the solution of Eq.~\eqref{eq:green3} with \mbox{$\varepsilon_\alpha=+ 1$ for all $\alpha$}. Indeed, the corresponding solution only holds up to a set of parameters beyond which one must consider a solution where one of the $\varepsilon_\alpha$ is negative. Equating the two corresponding Eq.~\eqref{eq:green3} immediately leads to the following condition:
%
%
\be 
J_z^2=J_x^2+J_y^2-2 \sqrt{(J_z^2-J_x^2)(J_z^2-J_y^2)}.
\label{eq:Kimchi2}
\ee 
%
%
This is in agreement with the results found in Ref.~\cite{Kimchi14} (see Appendix D). The corresponding lines shown in Fig.~\ref{fig:diagram} clearly illustrate the inability of the solution \mbox{$\varepsilon_\alpha=+ 1$ for all $\alpha$}, to determine the exact boundaries of the gapless phase. %


%
%
\section{Conclusions}
\label{sec:conclusion}
%
%
In this work, we examined the zero-temperature phase diagram of the Kitaev model defined on $\{p,3\}$ hyperbolic tilings. For finite $p>6$, we computed the fermion gap using ED on finite-size clusters embedded in closed surfaces to avoid boundary effects. We focused on two families of tilings that can accommodate either Kekul\'e ($p \mod 2=0$) or Kitaev ($p \mod 3=0$) three-edge colorings. In the former case, the gapless phase behaves like a bubble that grows as $p$ increases (from a single point at $p=6$ to an amoeba-like shape at $p=\infty$). In contrast, in the latter case, the gapless phase evolves from a triangular shape at $p=6$ to its asymptotic form at \mbox{$p=\infty$}, which corresponds to the trivalent Bethe lattice.  
We also obtained the exact boundaries of the phase diagram for the trivalent Bethe lattice by revisiting (and correcting) the results given in Ref.~\cite{Kimchi14}. In this limiting case, the phase diagram is independent of the coloring and, since there are no loops, it does not depend on the flux map. When one of the couplings dominates (isolated-dimer limit), we derived the low-energy effective Hamiltonians that describe a toric code phase. 

An interesting open question concerns the nature of the gapless phase. For $p=6$ (honeycomb lattice), Kitaev showed that this phase hosts non-Abelian (Ising) anyons when gapped by a three-spin term~\cite{Kitaev06}. Interestingly, this is also the case for $p=8$~\cite{Lenggenhager25} and $p=9$~\cite{Dusel25}, for which a Chern number $\nu=\pm 1$ has been found around the isotropic point. It would be nice to investigate this problem for other values of $p$, including $p=\infty$. 
Finally, we note that our focus here was on regular colorings. Considering random (three-edge) colorings could lead to new interesting physics related to the recently studied Anderson localization in hyperbolic lattices~\cite{Li24,Chen24}. 
 
\acknowledgments
 We thank  J.-N. Fuchs and O. Tchernyshyov for fruitful discussions. R.M. also thanks J.-P. Gaspard for early collaboration about the spectrum of Bethe lattice of polygons.

\appendix
\section{Trivalent three-coloured Bethe lattice}
\label{app:Bethe}

To compute the  density of states of the trivalent three-colored Bethe lattice, we consider the Green's function 
%
%
\be
G(E)=\frac{1}{E-H},
\ee
%
%
which satisfies the Dyson equation
%
%
\be
G(E)=\frac{1}{E}+\frac{1}{E} \: H \: G(E).
\ee
%
Then, we analyze the resulting hierarchy of equations starting from the $G_{00}$ matrix element where $0$ denotes an (arbitrary) given site $0$. We apply the standard method for tree-like structures, which amounts to studying an effective Hamiltonian limited to a central cluster dressed with energy-dependent diagonal self-energies to take account for all neglected contributions. This procedure is known to be exact for tree graphs, provided that self-consistent equations are considered for the self-energies. This ensures that the local Green's function is indeed site independent. When only the central site $0$ is kept, this leads to (see Appendix D in Ref.~\cite{Kimchi14})
%
%
\be
G_{00}=\frac{1}{E-s_x-s_y-s_z}= \frac{1}{E-S},
\label{eq:green0}
\ee
%
%
where $s_\alpha$ is the self-energy associated with couplings $J_\alpha$, and where $S=s_x+s_y+s_z$. From here on, we will omit the dependence with $E$ of the Green's function.

%
%
\begin{figure}[t]
\includegraphics[width=.5\columnwidth ]{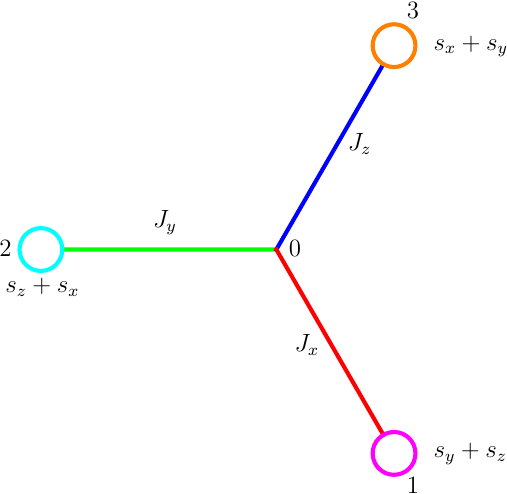}
\caption{Local cluster of the trivalent Bethe lattice. Links are associated with hopping terms whereas circles represent self-energies (see text). }
\label{fig:selfbethe}
\end{figure} 
%
%
If we consider a larger cluster consisting of a central site and its three neighbors (see Fig.~\ref{fig:selfbethe}), one gets
%
%
\be
E \: G_{00}=1+J_x G_{10}+ J_y G_{20}+J_z G_{30},
\label{eq:first}
\ee
%
%
and
%
%
\beqn
(E-s_y-s_z) G_{10}&=& J_x G_{00},\\
(E-s_z-s_x) G_{20}&=& J_y G_{00},\\
(E-s_x-s_y) G_{30}&=& J_z G_{00},
\label{eq:last}
\eeqn
%
%
which yields:
%
%
\be
G_{00}=\frac{1}{E-\frac{J_x^2}{E-S+s_x} -\frac{J_y^2}{E-S+s_y}-\frac{J_z^2}{E-S+s_z}}.
\label{eq:G00bis}
\ee
%
%
Note that, in Eqs.~\eqref{eq:first}-\eqref{eq:last}, we consider  the matrix elements of ${\rm i} A/2$ with $J_\alpha$'s as hopping terms.
Equating \eqref{eq:green0} and \eqref{eq:G00bis} leads to the self-consistent equation,
%
%
\beqn
s_\alpha= \frac{J_\alpha^2}{E-S+s_\alpha}.
\eeqn
%
%
The three self-energies $s_\alpha$  are then given by one of the two solutions,
%
\be
s_\alpha=\frac{-E+S +\varepsilon_\alpha \sqrt{(E-S)^2+4 J_\alpha^2}}{2},
\label{eq:sign}
\ee
%
%
with $\varepsilon_\alpha=\pm 1$. By summing over Eq.~\eqref{eq:sign} over \mbox{$\alpha=x,y,z$}, one obtains

\be
S=3E- \sum_{\alpha=x,y,z} \varepsilon_\alpha \sqrt{(E-S)^2+4 J_\alpha^2},
\label{eq:S}
\ee

Using Eq.~\eqref{eq:green0}, one then gets the following expression for the local Green's function:

%
%
\be
G_{00}^{-1} = -2E+ \sum_{\alpha=x,y,z} \varepsilon_\alpha \sqrt{G_{00}^{-2}+4 J_\alpha^2}.
\label{eq:green}
\ee
%
%
 Recall that, since all sites are equivalent, the above equation applies to any $G_{ii}$. In Appendix D of Ref.~\cite{Kimchi14}, the authors only consider the solution to Eq.~\eqref{eq:green} corresponding to $\varepsilon_\alpha=+ 1$. However, a close inspection of the entire set of equations reveals that other solutions are also relevant.

Indeed, determining the region where the gap vanishes at $E=0$ is equivalent to finding the set of parameters $J_\alpha$ for which the local density of states at site $i$ vanishes, i.e., $\rho_i(E=0)=-\frac{1}{\pi} {\rm Im}G_{ii}(0) =0$. As discussed in  Ref.~\cite{Kimchi14}, this never happens with the choice \mbox{$\varepsilon_\alpha=+ 1$, for all $\alpha$}. Nevertheless, this is possible with other sign choices as we shall now discuss. Indeed, setting $E=0$ and assuming $\rho_i(0)\neq 0$ (gapless phase), Eq.~\eqref{eq:green} can be recasted as
%
%
\be
1= \sum_{\alpha=x,y,z} \varepsilon_\alpha \sqrt{1-J_\alpha^2 r^2},
\label{eq:green2}
\ee
%
%
 where, following Appendix D in Ref.~\cite{Kimchi14}, we introduced $r=2\pi \rho_i (0)$. Here, we also used the fact that ${\rm Re}\: \:G_{ii}(0)=0$, which is always the case for a Majorana-fermion spectrum since $\rho_i(E)=\rho_i(-E)$.
 
Equation~\eqref{eq:green2} allows one to obtain the exact boundary between gapped and gapless phases as explained in Sec.~\ref{sec:Bethe}.
\section{Kitaev model for a Bethe lattice of polygons}
\label{app:Hecke}
%

\begin{figure}[t]
\includegraphics[width=0.493\columnwidth ]{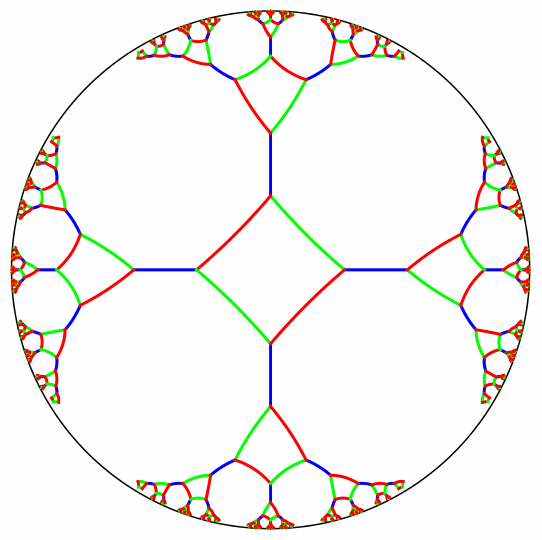}
\includegraphics[width=0.493\columnwidth ]{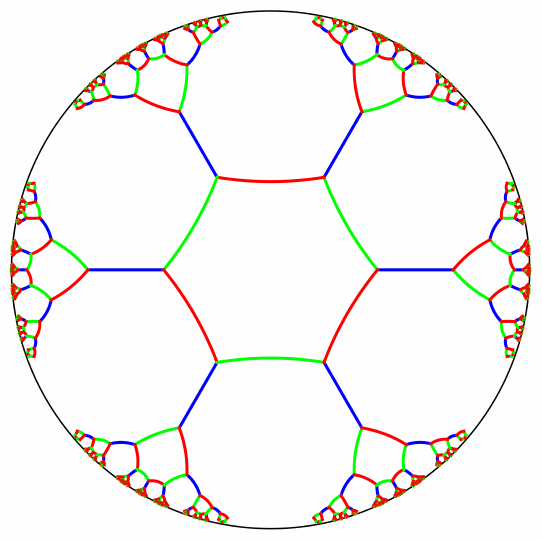}

\caption{Pieces of $\mathcal{H}_4$ (left) and $\mathcal{H}_6$ (right). Each site of $\mathcal{H}_p$ is trivalent and belongs to one $p$-gon and two apeirogons (infinite-length polygons). The limit $p\rightarrow \infty$ corresponds to the trivalent Bethe lattice (see Fig.~\ref{fig:disk_8}) for which the coloring is irrelevant.}
\label{fig:Hecke}
\end{figure}

In the main text of this paper, we analyzed how the Kitaev model defined on $\{p,3\}$ tilings and we explained how the phase diagram converges to the one of the Bethe lattice as $p$ goes to infinity. In this appendix, we propose an alternative route to reach this limit. We consider a Bethe lattice of polygons $\mathcal{H}_p$ obtained by replacing each site of the usual $p$-valent Bethe lattice by a $p$-gon as illustrated in Fig.~\ref{fig:Hecke} for $p=4$ (left)  and $p=6$ (right). The name  $\mathcal{H}_p$ comes from its relation with the Cayley graph for Hecke group $\mathbb{Z}_2 \times \mathbb{Z}_p$~\cite{Mairesse07}. However, contrary to the Cayley graph whose edges correspond to the group generator' action, we focus on a Kekul\'e coloring (only valid for even $p$), which allows us to study the Kitaev model case. In the following, we consider the case where each $p$-gon, related by $z$ links, consists of alternating  $x$ and $y$ links.  As such, the phase diagram of the Kitaev model for $\mathcal{H}_p$ is not invariant under the permutation of $J_x$, $J_y$, and $J_z$ coupling constants.

%
%
%
%

%
%
\begin{figure}[b]
\includegraphics[width=0.5\columnwidth ]{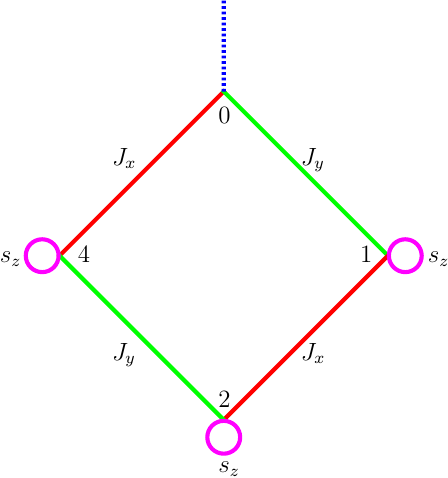}
\caption{A piece of  $\mathcal{H}_4$ used to compute  $g_{00}$. The outgoing edge at site $0$ is missing, while other outgoing edges are replaced by self-energies $s_z$.}
\label{fig:clusterHecke}
\end{figure} 
%
%

To compute the Green's function, we proceed in two steps, which have already been described for a different, but closely related set, the Husimi cactus of polygons~\cite{Thorpe73}. We first obtain the local Green's function $g_{00}$, corresponding to an isolated $p$-gon, with one edge missing at the site $0$, where the local density of states is computed, and all other $(p-1)$ sites being dressed by a self-energy $s_z$ accounting for the missing contributions through the outgoing edges (see Fig.~\ref{fig:clusterHecke}). 

Once $g_{00}$ is computed, the full Green's function $G_{00}$ reads
%
%
\be
G_{00}=\frac{1}{g_{00}^{-1}-s_z}= \frac{1}{g_{00}^{-1}-J_z^2 \:g_{00}},
\label{eq:g00}
 \ee
%
%
the latter formulation coming from the self-consistency relation, $s_z=J_z^2 \:g_{00}$ (recall that $g_{00}$ depends on $s_z$ via the self-energies $s_z$).
 
%
%
\begin{figure}[t]
\includegraphics[width=0.8\columnwidth ]{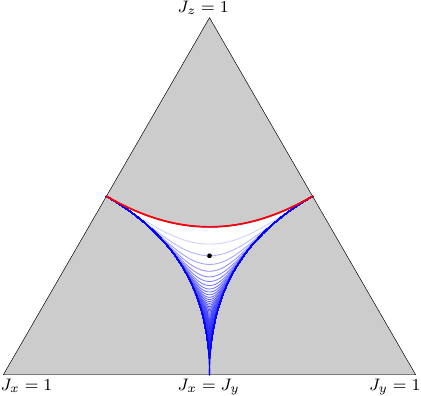}
\caption{Phase diagram of $\mathcal{H}_p$ for different values of $p$. For each value of $p$, the gapless phase is delimited by the red line $J_z^2=J_x^2+J_y^2$, and by the corresponding blue line given by Eqs.~\eqref{eq:bound1} and \eqref{eq:bound2} (drawn for $p=4,6,8,\dots$, from top to bottom). Interestingly, for $p=4$, there is no gapless phase but just a gapless line separating two gapped phases (the blue and the red lines are the same  in this case). For $p=8$, the blue line contains the isotropic point (black dot). The Bethe lattice limit is obtained as $p\rightarrow \infty$.} 
\label{fig:Heckephasediagram}
\end{figure} 
%
We first compute $g_{00}$ as a finite continued fraction~\cite{Haydock72,Gaspard73,Haydock75},
 
%
%
\be
g_{00}=\frac{1}{E-\frac{b_1}{E-s_z-\frac{b_2}{E-s_z - ...}}}.
\label{eq:fraccont}
\ee
%
%
For a $p$-gon consisting of alternating $x$ and $y$ links, the $p/2$ coefficients $b_i$ can be computed exactly. They are given by
%
%
\beqn
b_1&=&J_x^2+J_y^2, \\
b_{2j}&=& J_x^2 J_y^2 \frac{J_x^{2(j-1)}+J_y^{2(j-1)}}{J_x^{2j}+J_y^{2j}},  \\
b_{2j+1}&=&\frac{J_x^{2(j+1)}+J_y^{2(j+1)}}{J_x^{2j}+J_y^{2j}}.
 \label{coeff}
 \eeqn
%
%
Because of the flux constraint given by the conjecture \eqref{eq:fluxmin}, the last coefficient is either $b_{p/2}=0$, if $p \mod 4=2$, or   
\be
b_{p/2}=\frac{J_x^{p/2+1}+J_x^{p/2} J_y+J_x J_y^{p/2}+J_y^{p/2+1}}{J_x^{p/2-1}+ J_x^{p/2-1}},
\ee
if $p \mod 4=0$.

Using these coefficients and Eqs.~\eqref{eq:g00} and \eqref{eq:fraccont}, one can easily find the conditions under which the system is gapless, [i.e., ${\rm Im}G_{00}(0) \neq 0$]. For all values of (even) $p$, one condition is given by 
\be
J_z^2=J_x^2+J_y^2,
\label{eq:bla}
\ee
(see red line in Fig.~\ref{fig:Heckephasediagram}). The other boundary is either given by
%
%
\be
J_z^2=\frac{J_x^{p/2+1}+J_x^{p/2} J_y+J_x J_y^{p/2}+J_y^{p/2+1}}{\sum_{n=1}^{p/2} J_x^{p/2-n}J_y^{n-1}},
\label{eq:bound1}
 \ee
%
%
if $p \mod 4=2$, or    
%
%
\be
J_z^2=\frac{J_x^{p/2}+J_y^{p/2}}{\sum_{n=1}^{p/4} J_x^{p/2-2n} J_y^{2n-2}},
\label{eq:bound2}
\ee
%
%
if $p \mod 4=0$ (see blue lines in Fig.~\ref{fig:Heckephasediagram}). Interestingly, for $p=4$, the corresponding line coincides with  Eq.~\eqref{eq:bla} so that there is no gapless phase but only a gapless line in this case.


\begin{thebibliography}{40}%
\makeatletter
\providecommand \@ifxundefined [1]{%
 \@ifx{#1\undefined}
}%
\providecommand \@ifnum [1]{%
 \ifnum #1\expandafter \@firstoftwo
 \else \expandafter \@secondoftwo
 \fi
}%
\providecommand \@ifx [1]{%
 \ifx #1\expandafter \@firstoftwo
 \else \expandafter \@secondoftwo
 \fi
}%
\providecommand \natexlab [1]{#1}%
\providecommand \enquote  [1]{``#1''}%
\providecommand \bibnamefont  [1]{#1}%
\providecommand \bibfnamefont [1]{#1}%
\providecommand \citenamefont [1]{#1}%
\providecommand \href@noop [0]{\@secondoftwo}%
\providecommand \href [0]{\begingroup \@sanitize@url \@href}%
\providecommand \@href[1]{\@@startlink{#1}\@@href}%
\providecommand \@@href[1]{\endgroup#1\@@endlink}%
\providecommand \@sanitize@url [0]{\catcode `\\12\catcode `\$12\catcode
  `\&12\catcode `\#12\catcode `\^12\catcode `\_12\catcode `\%12\relax}%
\providecommand \@@startlink[1]{}%
\providecommand \@@endlink[0]{}%
\providecommand \url  [0]{\begingroup\@sanitize@url \@url }%
\providecommand \@url [1]{\endgroup\@href {#1}{\urlprefix }}%
\providecommand \urlprefix  [0]{URL }%
\providecommand \Eprint [0]{\href }%
\providecommand \doibase [0]{https://doi.org/}%
\providecommand \selectlanguage [0]{\@gobble}%
\providecommand \bibinfo  [0]{\@secondoftwo}%
\providecommand \bibfield  [0]{\@secondoftwo}%
\providecommand \translation [1]{[#1]}%
\providecommand \BibitemOpen [0]{}%
\providecommand \bibitemStop [0]{}%
\providecommand \bibitemNoStop [0]{.\EOS\space}%
\providecommand \EOS [0]{\spacefactor3000\relax}%
\providecommand \BibitemShut  [1]{\csname bibitem#1\endcsname}%
\let\auto@bib@innerbib\@empty
\bibitem [{\citenamefont {Tsui}\ \emph {et~al.}(1982)\citenamefont {Tsui},
  \citenamefont {Stormer},\ and\ \citenamefont {Gossard}}]{Tsui82}%
  \BibitemOpen
  \bibfield  {author} {\bibinfo {author} {\bibfnamefont {D.~C.}\ \bibnamefont
  {Tsui}}, \bibinfo {author} {\bibfnamefont {H.~L.}\ \bibnamefont {Stormer}},\
  and\ \bibinfo {author} {\bibfnamefont {A.~C.}\ \bibnamefont {Gossard}},\
  }\bibfield  {title} {\bibinfo {title} {{Two-Dimensional Magnetotransport in
  the Extreme Quantum Limit}},\ }\href
  {https://doi.org/10.1103/PhysRevLett.48.1559} {\bibfield  {journal} {\bibinfo
   {journal} {Phys. Rev. Lett.}\ }\textbf {\bibinfo {volume} {48}},\ \bibinfo
  {pages} {1559} (\bibinfo {year} {1982})}\BibitemShut {NoStop}%
\bibitem [{\citenamefont {Stormer}\ \emph {et~al.}(1999)\citenamefont
  {Stormer}, \citenamefont {Tsui},\ and\ \citenamefont {Gossard}}]{Stormer99}%
  \BibitemOpen
  \bibfield  {author} {\bibinfo {author} {\bibfnamefont {H.~L.}\ \bibnamefont
  {Stormer}}, \bibinfo {author} {\bibfnamefont {T.~C.}\ \bibnamefont {Tsui}},\
  and\ \bibinfo {author} {\bibfnamefont {A.~C.}\ \bibnamefont {Gossard}},\
  }\bibfield  {title} {\bibinfo {title} {{The fractional quantum Hall
  effect}},\ }\href {https://doi.org/10.1103/RevModPhys.71.S298} {\bibfield
  {journal} {\bibinfo  {journal} {Rev. Mod. Phys.}\ }\textbf {\bibinfo {volume}
  {71}},\ \bibinfo {pages} {S298} (\bibinfo {year} {1999})}\BibitemShut
  {NoStop}%
\bibitem [{\citenamefont {Leinaas}\ and\ \citenamefont
  {Myrheim}(1977)}]{Leinaas77}%
  \BibitemOpen
  \bibfield  {author} {\bibinfo {author} {\bibfnamefont {J.~M.}\ \bibnamefont
  {Leinaas}}\ and\ \bibinfo {author} {\bibfnamefont {J.}~\bibnamefont
  {Myrheim}},\ }\bibfield  {title} {\bibinfo {title} {{On the theory of
  identical particles}},\ }\href {https://doi.org/10.1007/BF02727953}
  {\bibfield  {journal} {\bibinfo  {journal} {Nuovo Cim. B}\ }\textbf {\bibinfo
  {volume} {37}},\ \bibinfo {pages} {1} (\bibinfo {year} {1977})}\BibitemShut
  {NoStop}%
\bibitem [{\citenamefont {Wilczek}(1982{\natexlab{a}})}]{Wilczek82_1}%
  \BibitemOpen
  \bibfield  {author} {\bibinfo {author} {\bibfnamefont {F.}~\bibnamefont
  {Wilczek}},\ }\bibfield  {title} {\bibinfo {title} {{Magnetic flux, angular
  momentum, and statistics}},\ }\href
  {https://doi.org/10.1103/PhysRevLett.48.1144} {\bibfield  {journal} {\bibinfo
   {journal} {Phys. Rev. Lett.}\ }\textbf {\bibinfo {volume} {48}},\ \bibinfo
  {pages} {1144} (\bibinfo {year} {1982}{\natexlab{a}})}\BibitemShut {NoStop}%
\bibitem [{\citenamefont {Wilczek}(1982{\natexlab{b}})}]{Wilczek82_2}%
  \BibitemOpen
  \bibfield  {author} {\bibinfo {author} {\bibfnamefont {F.}~\bibnamefont
  {Wilczek}},\ }\bibfield  {title} {\bibinfo {title} {{Quantum mechanics of
  fractional-spin particles}},\ }\href
  {https://doi.org/10.1103/PhysRevLett.49.957} {\bibfield  {journal} {\bibinfo
  {journal} {Phys. Rev. Lett.}\ }\textbf {\bibinfo {volume} {49}},\ \bibinfo
  {pages} {957} (\bibinfo {year} {1982}{\natexlab{b}})}\BibitemShut {NoStop}%
\bibitem [{\citenamefont {Goldin}\ \emph {et~al.}(1985)\citenamefont {Goldin},
  \citenamefont {Menikoff},\ and\ \citenamefont {Sharp}}]{Goldin85}%
  \BibitemOpen
  \bibfield  {author} {\bibinfo {author} {\bibfnamefont {G.~A.}\ \bibnamefont
  {Goldin}}, \bibinfo {author} {\bibfnamefont {R.}~\bibnamefont {Menikoff}},\
  and\ \bibinfo {author} {\bibfnamefont {D.~H.}\ \bibnamefont {Sharp}},\
  }\bibfield  {title} {\bibinfo {title} {{Comments on "General theory for
  quantum statistics in two dimensions"}},\ }\href
  {https://doi.org/10.1103/PhysRevLett.54.603} {\bibfield  {journal} {\bibinfo
  {journal} {Phys. Rev. Lett.}\ }\textbf {\bibinfo {volume} {54}},\ \bibinfo
  {pages} {603} (\bibinfo {year} {1985})}\BibitemShut {NoStop}%
\bibitem [{\citenamefont {Bartolomei}\ \emph {et~al.}(2020)\citenamefont
  {Bartolomei}, \citenamefont {Kumar}, \citenamefont {Bisognin}, \citenamefont
  {Marguerite}, \citenamefont {Berroir}, \citenamefont {Bocquillon},
  \citenamefont {Pla\c{c}ais}, \citenamefont {Cavanna}, \citenamefont {Dong},
  \citenamefont {Gennser}, \citenamefont {Jin},\ and\ \citenamefont
  {F\`eve}}]{Bartolomei20}%
  \BibitemOpen
  \bibfield  {author} {\bibinfo {author} {\bibfnamefont {H.}~\bibnamefont
  {Bartolomei}}, \bibinfo {author} {\bibfnamefont {M.}~\bibnamefont {Kumar}},
  \bibinfo {author} {\bibfnamefont {R.}~\bibnamefont {Bisognin}}, \bibinfo
  {author} {\bibfnamefont {A.}~\bibnamefont {Marguerite}}, \bibinfo {author}
  {\bibfnamefont {J.-M.}\ \bibnamefont {Berroir}}, \bibinfo {author}
  {\bibfnamefont {E.}~\bibnamefont {Bocquillon}}, \bibinfo {author}
  {\bibfnamefont {B.}~\bibnamefont {Pla\c{c}ais}}, \bibinfo {author}
  {\bibfnamefont {A.}~\bibnamefont {Cavanna}}, \bibinfo {author} {\bibfnamefont
  {Q.}~\bibnamefont {Dong}}, \bibinfo {author} {\bibfnamefont {U.}~\bibnamefont
  {Gennser}}, \bibinfo {author} {\bibfnamefont {Y.}~\bibnamefont {Jin}},\ and\
  \bibinfo {author} {\bibfnamefont {G.}~\bibnamefont {F\`eve}},\ }\bibfield
  {title} {\bibinfo {title} {{Fractional statistics in anyon collisions}},\
  }\href {https://doi.org/10.1126/science.aaz5601} {\bibfield  {journal}
  {\bibinfo  {journal} {Science}\ }\textbf {\bibinfo {volume} {368}},\ \bibinfo
  {pages} {173} (\bibinfo {year} {2020})}\BibitemShut {NoStop}%
\bibitem [{\citenamefont {Nakamura}\ \emph {et~al.}(2020)\citenamefont
  {Nakamura}, \citenamefont {Liang}, \citenamefont {Gardner},\ and\
  \citenamefont {Manfra}}]{Nakamura20}%
  \BibitemOpen
  \bibfield  {author} {\bibinfo {author} {\bibfnamefont {J.}~\bibnamefont
  {Nakamura}}, \bibinfo {author} {\bibfnamefont {S.}~\bibnamefont {Liang}},
  \bibinfo {author} {\bibfnamefont {G.~C.}\ \bibnamefont {Gardner}},\ and\
  \bibinfo {author} {\bibfnamefont {M.~J.}\ \bibnamefont {Manfra}},\ }\bibfield
   {title} {\bibinfo {title} {{Direct observation of anyonic braiding
  statistics}},\ }\href {https://doi.org/10.1038/s41567-020-1019-1} {\bibfield
  {journal} {\bibinfo  {journal} {Nat. Phys.}\ }\textbf {\bibinfo {volume}
  {16}},\ \bibinfo {pages} {931} (\bibinfo {year} {2020})}\BibitemShut
  {NoStop}%
\bibitem [{\citenamefont {Wen}(2017)}]{Wen17}%
  \BibitemOpen
  \bibfield  {author} {\bibinfo {author} {\bibfnamefont {X.-G.}\ \bibnamefont
  {Wen}},\ }\bibfield  {title} {\bibinfo {title} {{Zoo of quantum-topological
  phases of matter}},\ }\href {https://doi.org/10.1103/RevModPhys.89.041004}
  {\bibfield  {journal} {\bibinfo  {journal} {Rev. Mod. Phys.}\ }\textbf
  {\bibinfo {volume} {89}},\ \bibinfo {pages} {041004} (\bibinfo {year}
  {2017})}\BibitemShut {NoStop}%
\bibitem [{\citenamefont {Nayak}\ \emph {et~al.}(2008)\citenamefont {Nayak},
  \citenamefont {Simon}, \citenamefont {Stern}, \citenamefont {Freedman},\ and\
  \citenamefont {Sarma}}]{Nayak08}%
  \BibitemOpen
  \bibfield  {author} {\bibinfo {author} {\bibfnamefont {C.}~\bibnamefont
  {Nayak}}, \bibinfo {author} {\bibfnamefont {S.~H.}\ \bibnamefont {Simon}},
  \bibinfo {author} {\bibfnamefont {A.}~\bibnamefont {Stern}}, \bibinfo
  {author} {\bibfnamefont {M.}~\bibnamefont {Freedman}},\ and\ \bibinfo
  {author} {\bibfnamefont {S.~D.}\ \bibnamefont {Sarma}},\ }\bibfield  {title}
  {\bibinfo {title} {{Non-Abelian anyons and topological quantum
  computation}},\ }\href {https://doi.org/10.1103/RevModPhys.80.1083}
  {\bibfield  {journal} {\bibinfo  {journal} {Rev. Mod. Phys.}\ }\textbf
  {\bibinfo {volume} {80}},\ \bibinfo {pages} {1083} (\bibinfo {year}
  {2008})}\BibitemShut {NoStop}%
\bibitem [{\citenamefont {Kitaev}(2006)}]{Kitaev06}%
  \BibitemOpen
  \bibfield  {author} {\bibinfo {author} {\bibfnamefont {A.}~\bibnamefont
  {Kitaev}},\ }\bibfield  {title} {\bibinfo {title} {Anyons in an exactly
  solved model and beyond},\ }\href {https://doi.org/10.1016/j.aop.2005.10.005}
  {\bibfield  {journal} {\bibinfo  {journal} {Ann. Phys. (NY)}\ }\textbf
  {\bibinfo {volume} {321}},\ \bibinfo {pages} {2} (\bibinfo {year}
  {2006})}\BibitemShut {NoStop}%
\bibitem [{\citenamefont {Trebst}\ and\ \citenamefont
  {Hickey}(2022)}]{Trebst22}%
  \BibitemOpen
  \bibfield  {author} {\bibinfo {author} {\bibfnamefont {S.}~\bibnamefont
  {Trebst}}\ and\ \bibinfo {author} {\bibfnamefont {C.}~\bibnamefont
  {Hickey}},\ }\bibfield  {title} {\bibinfo {title} {{Kitaev materials}},\
  }\href {https://doi.org/10.1016/j.physrep.2021.11.003} {\bibfield  {journal}
  {\bibinfo  {journal} {Phys. Rep.}\ }\textbf {\bibinfo {volume} {950}},\
  \bibinfo {pages} {1} (\bibinfo {year} {2022})}\BibitemShut {NoStop}%
\bibitem [{\citenamefont {Matsuda}\ \emph {et~al.}()\citenamefont {Matsuda},
  \citenamefont {Shibauchi},\ and\ \citenamefont {Kee}}]{Matsuda25}%
  \BibitemOpen
  \bibfield  {author} {\bibinfo {author} {\bibfnamefont {Y.}~\bibnamefont
  {Matsuda}}, \bibinfo {author} {\bibfnamefont {T.}~\bibnamefont {Shibauchi}},\
  and\ \bibinfo {author} {\bibfnamefont {H.-Y.}\ \bibnamefont {Kee}},\
  }\href@noop {} {}\bibinfo {note} {Kitaev quantum spin liquids,
  \href{https://arxiv.org/abs/2501.05608}{arXiv:2501.05608}}\BibitemShut
  {NoStop}%
\bibitem [{\citenamefont {Mosseri}\ \emph {et~al.}(2025)\citenamefont
  {Mosseri}, \citenamefont {Iqbal}, \citenamefont {Vogeler},\ and\
  \citenamefont {Vidal}}]{Mosseri25}%
  \BibitemOpen
  \bibfield  {author} {\bibinfo {author} {\bibfnamefont {R.}~\bibnamefont
  {Mosseri}}, \bibinfo {author} {\bibfnamefont {Y.}~\bibnamefont {Iqbal}},
  \bibinfo {author} {\bibfnamefont {R.}~\bibnamefont {Vogeler}},\ and\ \bibinfo
  {author} {\bibfnamefont {J.}~\bibnamefont {Vidal}},\ }\bibfield  {title}
  {\bibinfo {title} {{Kitaev model on Hurwitz hyperbolic tilings: A non-Abelian
  gapped chiral spin liquid}},\ }\href
  {https://doi.org/10.1103/PhysRevB.111.L060408} {\bibfield  {journal}
  {\bibinfo  {journal} {Phys. Rev. B}\ }\textbf {\bibinfo {volume} {111}},\
  \bibinfo {pages} {L060408} (\bibinfo {year} {2025})}\BibitemShut {NoStop}%
\bibitem [{\citenamefont {Lenggenhager}\ \emph {et~al.}(2025)\citenamefont
  {Lenggenhager}, \citenamefont {Dey}, \citenamefont {Bzdu\v{s}ek},\ and\
  \citenamefont {Maciejko}}]{Lenggenhager25}%
  \BibitemOpen
  \bibfield  {author} {\bibinfo {author} {\bibfnamefont {P.~M.}\ \bibnamefont
  {Lenggenhager}}, \bibinfo {author} {\bibfnamefont {S.}~\bibnamefont {Dey}},
  \bibinfo {author} {\bibfnamefont {T.}~\bibnamefont {Bzdu\v{s}ek}},\ and\
  \bibinfo {author} {\bibfnamefont {J.}~\bibnamefont {Maciejko}},\ }\bibfield
  {title} {\bibinfo {title} {{Hyperbolic spin liquids}},\ }\href
  {https://doi.org/https://doi.org/10.1103/s25y-s4fj} {\bibfield  {journal}
  {\bibinfo  {journal} {Phys. Rev. Lett.}\ }\textbf {\bibinfo {volume} {135}},\
  \bibinfo {pages} {076604} (\bibinfo {year} {2025})}\BibitemShut {NoStop}%
\bibitem [{\citenamefont {Dusel}\ \emph {et~al.}(2025)\citenamefont {Dusel},
  \citenamefont {Hofmann}, \citenamefont {Maity}, \citenamefont {Mosseri},
  \citenamefont {Vidal}, \citenamefont {Iqbal}, \citenamefont {Greiter},\ and\
  \citenamefont {Thomale}}]{Dusel25}%
  \BibitemOpen
  \bibfield  {author} {\bibinfo {author} {\bibfnamefont {F.}~\bibnamefont
  {Dusel}}, \bibinfo {author} {\bibfnamefont {T.}~\bibnamefont {Hofmann}},
  \bibinfo {author} {\bibfnamefont {A.}~\bibnamefont {Maity}}, \bibinfo
  {author} {\bibfnamefont {R.}~\bibnamefont {Mosseri}}, \bibinfo {author}
  {\bibfnamefont {J.}~\bibnamefont {Vidal}}, \bibinfo {author} {\bibfnamefont
  {Y.}~\bibnamefont {Iqbal}}, \bibinfo {author} {\bibfnamefont
  {M.}~\bibnamefont {Greiter}},\ and\ \bibinfo {author} {\bibfnamefont
  {R.}~\bibnamefont {Thomale}},\ }\bibfield  {title} {\bibinfo {title} {Chiral
  gapless spin liquid in hyperbolic space},\ }\href
  {https://doi.org/https://doi.org/10.1103/PhysRevLett.134.256604} {\bibfield
  {journal} {\bibinfo  {journal} {Phys. Rev. Lett.}\ }\textbf {\bibinfo
  {volume} {134}},\ \bibinfo {pages} {256604} (\bibinfo {year}
  {2025})}\BibitemShut {NoStop}%
\bibitem [{\citenamefont {Mosseri}\ and\ \citenamefont
  {Sadoc}(1982)}]{Mosseri82}%
  \BibitemOpen
  \bibfield  {author} {\bibinfo {author} {\bibfnamefont {R.}~\bibnamefont
  {Mosseri}}\ and\ \bibinfo {author} {\bibfnamefont {J.~F.}\ \bibnamefont
  {Sadoc}},\ }\bibfield  {title} {\bibinfo {title} {{The Bethe lattice: a
  regular tiling of the hyperbolic plane}},\ }\href
  {https://doi.org/10.1051/jphyslet:01982004308024900} {\bibfield  {journal}
  {\bibinfo  {journal} {J. Phys. Lett.}\ }\textbf {\bibinfo {volume} {43}},\
  \bibinfo {pages} {249} (\bibinfo {year} {1982})}\BibitemShut {NoStop}%
\bibitem [{\citenamefont {Kimchi}\ \emph {et~al.}(2014)\citenamefont {Kimchi},
  \citenamefont {Analytis},\ and\ \citenamefont {Vishwanath}}]{Kimchi14}%
  \BibitemOpen
  \bibfield  {author} {\bibinfo {author} {\bibfnamefont {I.}~\bibnamefont
  {Kimchi}}, \bibinfo {author} {\bibfnamefont {J.~G.}\ \bibnamefont
  {Analytis}},\ and\ \bibinfo {author} {\bibfnamefont {A.}~\bibnamefont
  {Vishwanath}},\ }\bibfield  {title} {\bibinfo {title} {{Three-dimensional
  quantum spin liquids in models of harmonic-honeycomb iridates and phase
  diagram in an infinite-$D$ approximation}},\ }\href
  {https://doi.org/10.1103/PhysRevB.90.205126} {\bibfield  {journal} {\bibinfo
  {journal} {Phys. Rev. B}\ }\textbf {\bibinfo {volume} {90}},\ \bibinfo
  {pages} {205126} (\bibinfo {year} {2014})}\BibitemShut {NoStop}%
\bibitem [{\citenamefont {Lieb}\ and\ \citenamefont {Loss}(1993)}]{Lieb93}%
  \BibitemOpen
  \bibfield  {author} {\bibinfo {author} {\bibfnamefont {E.~H.}\ \bibnamefont
  {Lieb}}\ and\ \bibinfo {author} {\bibfnamefont {M.}~\bibnamefont {Loss}},\
  }\bibfield  {title} {\bibinfo {title} {{Fluxes, Laplacians, and Kasteleyn's
  theorem}},\ }\href {https://doi.org/10.1215/S0012-7094-93-07114-1} {\bibfield
   {journal} {\bibinfo  {journal} {Duke Math. J.}\ }\textbf {\bibinfo {volume}
  {71}},\ \bibinfo {pages} {337} (\bibinfo {year} {1993})}\BibitemShut
  {NoStop}%
\bibitem [{\citenamefont {{G. Cassella and P. d'Ornellas and T. Hodson and W.
  M. H. Natori and J. Knolle}}(2023)}]{Cassella23}%
  \BibitemOpen
  \bibfield  {author} {\bibinfo {author} {\bibnamefont {{G. Cassella and P.
  d'Ornellas and T. Hodson and W. M. H. Natori and J. Knolle}}},\ }\bibfield
  {title} {\bibinfo {title} {{An exact chiral amorphous spin liquid}},\ }\href
  {https://doi.org/10.1038/s41467-023-42105-9} {\bibfield  {journal} {\bibinfo
  {journal} {Nat. Commun.}\ }\textbf {\bibinfo {volume} {14}},\ \bibinfo
  {pages} {6663} (\bibinfo {year} {2023})}\BibitemShut {NoStop}%
\bibitem [{\citenamefont {Petrova}\ \emph {et~al.}(2014)\citenamefont
  {Petrova}, \citenamefont {Mellado},\ and\ \citenamefont
  {Tchernyshyov}}]{Petrova14}%
  \BibitemOpen
  \bibfield  {author} {\bibinfo {author} {\bibfnamefont {O.}~\bibnamefont
  {Petrova}}, \bibinfo {author} {\bibfnamefont {P.}~\bibnamefont {Mellado}},\
  and\ \bibinfo {author} {\bibfnamefont {O.}~\bibnamefont {Tchernyshyov}},\
  }\bibfield  {title} {\bibinfo {title} {{Unpaired Majorana modes on
  dislocations and string defects in Kitaev's honeycomb model}},\ }\href
  {https://doi.org/10.1103/PhysRevB.90.134404} {\bibfield  {journal} {\bibinfo
  {journal} {Phys. Rev. B}\ }\textbf {\bibinfo {volume} {90}},\ \bibinfo
  {pages} {134404} (\bibinfo {year} {2014})}\BibitemShut {NoStop}%
\bibitem [{\citenamefont {Kramers}(1930)}]{Kramers31}%
  \BibitemOpen
  \bibfield  {author} {\bibinfo {author} {\bibfnamefont {H.~A.}\ \bibnamefont
  {Kramers}},\ }\bibfield  {title} {\bibinfo {title} {{Th\'eorie g\'en\'erale
  de la rotation paramagn\'etique dans les cristaux}},\ }\href
  {https://doi.org/https://dwc.knaw.nl/DL/publications/PU00015981.pdf}
  {\bibfield  {journal} {\bibinfo  {journal} {Proc. Amsterdam Acad.}\ }\textbf
  {\bibinfo {volume} {33}},\ \bibinfo {pages} {959} (\bibinfo {year}
  {1930})}\BibitemShut {NoStop}%
\bibitem [{\citenamefont {Yao}\ and\ \citenamefont {Kivelson}(2007)}]{Yao07}%
  \BibitemOpen
  \bibfield  {author} {\bibinfo {author} {\bibfnamefont {H.}~\bibnamefont
  {Yao}}\ and\ \bibinfo {author} {\bibfnamefont {S.~A.}\ \bibnamefont
  {Kivelson}},\ }\bibfield  {title} {\bibinfo {title} {{Exact chiral spin
  liquid with non-Abelian anyons}},\ }\href
  {https://doi.org/10.1103/PhysRevLett.99.247203} {\bibfield  {journal}
  {\bibinfo  {journal} {Phys. Rev. Lett.}\ }\textbf {\bibinfo {volume} {99}},\
  \bibinfo {pages} {247203} (\bibinfo {year} {2007})}\BibitemShut {NoStop}%
\bibitem [{\citenamefont {Dusuel}\ \emph {et~al.}(2008)\citenamefont {Dusuel},
  \citenamefont {Schmidt}, \citenamefont {Vidal},\ and\ \citenamefont
  {Zaffino}}]{Dusuel08_2}%
  \BibitemOpen
  \bibfield  {author} {\bibinfo {author} {\bibfnamefont {S.}~\bibnamefont
  {Dusuel}}, \bibinfo {author} {\bibfnamefont {K.~P.}\ \bibnamefont {Schmidt}},
  \bibinfo {author} {\bibfnamefont {J.}~\bibnamefont {Vidal}},\ and\ \bibinfo
  {author} {\bibfnamefont {R.~L.}\ \bibnamefont {Zaffino}},\ }\bibfield
  {title} {\bibinfo {title} {{Perturbative study of the Kitaev model with
  spontaneous time-reversal symmetry breaking}},\ }\href
  {https://doi.org/10.1103/PhysRevB.78.125102} {\bibfield  {journal} {\bibinfo
  {journal} {Phys. Rev. B}\ }\textbf {\bibinfo {volume} {78}},\ \bibinfo
  {pages} {125102} (\bibinfo {year} {2008})}\BibitemShut {NoStop}%
\bibitem [{Kam()}]{Kamfor10}%
  \BibitemOpen
  \href@noop {} {}\bibinfo {note} {M. Kamfor, S. Dusuel, J. Vidal, and K. P.
  Schmidt, Kitaev model and dimer coverings on the honeycomb lattice,
  \href{http://iopscience.iop.org/1742-5468/2010/08/P08010/}{J. Stat. Mech.
  P08010 (2010)}}\BibitemShut {NoStop}%
\bibitem [{\citenamefont {Grushin}\ and\ \citenamefont
  {Repellin}(2023)}]{Grushin23}%
  \BibitemOpen
  \bibfield  {author} {\bibinfo {author} {\bibfnamefont {A.~G.}\ \bibnamefont
  {Grushin}}\ and\ \bibinfo {author} {\bibfnamefont {C.}~\bibnamefont
  {Repellin}},\ }\bibfield  {title} {\bibinfo {title} {{Amorphous and
  polycrystalline routes towards a chiral spin liquid}},\ }\href
  {https://doi.org/10.1103/PhysRevLett.130.186702} {\bibfield  {journal}
  {\bibinfo  {journal} {Phys. Rev. Lett.}\ }\textbf {\bibinfo {volume} {130}},\
  \bibinfo {pages} {186702} (\bibinfo {year} {2023})}\BibitemShut {NoStop}%
\bibitem [{\citenamefont {{R. Mosseri and J. Vidal}}(2023)}]{Mosseri23}%
  \BibitemOpen
  \bibfield  {author} {\bibinfo {author} {\bibnamefont {{R. Mosseri and J.
  Vidal}}},\ }\bibfield  {title} {\bibinfo {title} {{Density of states of
  tight-binding models in the hyperbolic plane}},\ }\href
  {https://doi.org/10.1103/PhysRevB.108.035154} {\bibfield  {journal} {\bibinfo
   {journal} {Phys. Rev. B}\ }\textbf {\bibinfo {volume} {108}},\ \bibinfo
  {pages} {035154} (\bibinfo {year} {2023})}\BibitemShut {NoStop}%
\bibitem [{\citenamefont {Conder}()}]{Conder06}%
  \BibitemOpen
  \bibfield  {author} {\bibinfo {author} {\bibfnamefont {M.}~\bibnamefont
  {Conder}},\ }\href@noop {} {\bibinfo {title} {Trivalent (cubic) symmetric
  graphs on up to 10000 vertices}},\ \bibinfo {note}
  {\href{https://www.math.auckland.ac.nz/~conder/}{https://www.math.auckland.ac.nz/~conder/}}\BibitemShut
  {NoStop}%
\bibitem [{\citenamefont {Pedrocchi}\ \emph {et~al.}(2011)\citenamefont
  {Pedrocchi}, \citenamefont {Chesi},\ and\ \citenamefont
  {Loss}}]{Pedrocchi11}%
  \BibitemOpen
  \bibfield  {author} {\bibinfo {author} {\bibfnamefont {F.~L.}\ \bibnamefont
  {Pedrocchi}}, \bibinfo {author} {\bibfnamefont {S.}~\bibnamefont {Chesi}},\
  and\ \bibinfo {author} {\bibfnamefont {D.}~\bibnamefont {Loss}},\ }\bibfield
  {title} {\bibinfo {title} {{Physical solutions of the Kitaev honeycomb
  model}},\ }\href {https://doi.org/10.1103/PhysRevB.84.165414} {\bibfield
  {journal} {\bibinfo  {journal} {Phys. Rev. B}\ }\textbf {\bibinfo {volume}
  {84}},\ \bibinfo {pages} {165414} (\bibinfo {year} {2011})}\BibitemShut
  {NoStop}%
\bibitem [{\citenamefont {Stegmaier}\ \emph {et~al.}(2022)\citenamefont
  {Stegmaier}, \citenamefont {Upreti}, \citenamefont {Thomale},\ and\
  \citenamefont {Boettcher}}]{Stegmaier22}%
  \BibitemOpen
  \bibfield  {author} {\bibinfo {author} {\bibfnamefont {A.}~\bibnamefont
  {Stegmaier}}, \bibinfo {author} {\bibfnamefont {L.~K.}\ \bibnamefont
  {Upreti}}, \bibinfo {author} {\bibfnamefont {R.}~\bibnamefont {Thomale}},\
  and\ \bibinfo {author} {\bibfnamefont {I.}~\bibnamefont {Boettcher}},\
  }\bibfield  {title} {\bibinfo {title} {{Universality of Hofstadter
  butterflies on hyperbolic lattices}},\ }\href
  {https://doi.org/10.1103/PhysRevLett.128.166402} {\bibfield  {journal}
  {\bibinfo  {journal} {Phys. Rev. Lett.}\ }\textbf {\bibinfo {volume} {128}},\
  \bibinfo {pages} {166402} (\bibinfo {year} {2022})}\BibitemShut {NoStop}%
\bibitem [{\citenamefont {Vidal}\ \emph {et~al.}(2008)\citenamefont {Vidal},
  \citenamefont {Schmidt},\ and\ \citenamefont {Dusuel}}]{Vidal08_2}%
  \BibitemOpen
  \bibfield  {author} {\bibinfo {author} {\bibfnamefont {J.}~\bibnamefont
  {Vidal}}, \bibinfo {author} {\bibfnamefont {K.~P.}\ \bibnamefont {Schmidt}},\
  and\ \bibinfo {author} {\bibfnamefont {S.}~\bibnamefont {Dusuel}},\
  }\bibfield  {title} {\bibinfo {title} {{Perturbative approach to an exactly
  solved problem: the Kitaev honeycomb model}},\ }\href
  {https://doi.org/10.1103/PhysRevB.78.245121} {\bibfield  {journal} {\bibinfo
  {journal} {Phys. Rev. B}\ }\textbf {\bibinfo {volume} {78}},\ \bibinfo
  {pages} {245121} (\bibinfo {year} {2008})}\BibitemShut {NoStop}%
\bibitem [{\citenamefont {Mandal}\ and\ \citenamefont
  {Surendran}(2009)}]{Mandal09}%
  \BibitemOpen
  \bibfield  {author} {\bibinfo {author} {\bibfnamefont {S.}~\bibnamefont
  {Mandal}}\ and\ \bibinfo {author} {\bibfnamefont {N.}~\bibnamefont
  {Surendran}},\ }\bibfield  {title} {\bibinfo {title} {{Exactly solvable
  Kitaev model in three dimensions}},\ }\href
  {https://doi.org/10.1103/PhysRevB.79.024426} {\bibfield  {journal} {\bibinfo
  {journal} {Phys. Rev. B}\ }\textbf {\bibinfo {volume} {79}},\ \bibinfo
  {pages} {024426} (\bibinfo {year} {2009})}\BibitemShut {NoStop}%
\bibitem [{\citenamefont {Pachos}(2007)}]{Pachos07}%
  \BibitemOpen
  \bibfield  {author} {\bibinfo {author} {\bibfnamefont {J.~K.}\ \bibnamefont
  {Pachos}},\ }\bibfield  {title} {\bibinfo {title} {The wavefunction of an
  anyon},\ }\href {https://doi.org/10.1016/j.aop.2006.05.007} {\bibfield
  {journal} {\bibinfo  {journal} {Ann. Phys. (NY)}\ }\textbf {\bibinfo {volume}
  {322}},\ \bibinfo {pages} {1254} (\bibinfo {year} {2007})}\BibitemShut
  {NoStop}%
\bibitem [{\citenamefont {Li}\ \emph {et~al.}(2024)\citenamefont {Li},
  \citenamefont {Peng}, \citenamefont {Wang},\ and\ \citenamefont {Hu}}]{Li24}%
  \BibitemOpen
  \bibfield  {author} {\bibinfo {author} {\bibfnamefont {T.}~\bibnamefont
  {Li}}, \bibinfo {author} {\bibfnamefont {Y.}~\bibnamefont {Peng}}, \bibinfo
  {author} {\bibfnamefont {Y.}~\bibnamefont {Wang}},\ and\ \bibinfo {author}
  {\bibfnamefont {H.}~\bibnamefont {Hu}},\ }\bibfield  {title} {\bibinfo
  {title} {{Anderson transition and mobility edges on hyperbolic lattices with
  randomly connected boundaries}},\ }\href
  {https://doi.org/10.1038/s42005-024-01848-7} {\bibfield  {journal} {\bibinfo
  {journal} {Commun. Phys.}\ }\textbf {\bibinfo {volume} {7}},\ \bibinfo
  {pages} {371} (\bibinfo {year} {2024})}\BibitemShut {NoStop}%
\bibitem [{\citenamefont {Chen}\ \emph {et~al.}(2024)\citenamefont {Chen},
  \citenamefont {Maciejko},\ and\ \citenamefont {Boettcher}}]{Chen24}%
  \BibitemOpen
  \bibfield  {author} {\bibinfo {author} {\bibfnamefont {A.}~\bibnamefont
  {Chen}}, \bibinfo {author} {\bibfnamefont {J.}~\bibnamefont {Maciejko}},\
  and\ \bibinfo {author} {\bibfnamefont {I.}~\bibnamefont {Boettcher}},\
  }\bibfield  {title} {\bibinfo {title} {{Anderson localization transition in
  disordered hyperbolic Lattices}},\ }\href
  {https://doi.org/10.1103/PhysRevLett.133.066101} {\bibfield  {journal}
  {\bibinfo  {journal} {Phys. Rev. Lett.}\ }\textbf {\bibinfo {volume} {133}},\
  \bibinfo {pages} {066101} (\bibinfo {year} {2024})}\BibitemShut {NoStop}%
\bibitem [{\citenamefont {Mairesse}\ and\ \citenamefont
  {Math\'eus}(2007)}]{Mairesse07}%
  \BibitemOpen
  \bibfield  {author} {\bibinfo {author} {\bibfnamefont {J.}~\bibnamefont
  {Mairesse}}\ and\ \bibinfo {author} {\bibfnamefont {F.}~\bibnamefont
  {Math\'eus}},\ }\bibfield  {title} {\bibinfo {title} {Random walks on free
  products of cyclic groups},\ }\href {https://doi.org/10.1112/jlms/jdl006}
  {\bibfield  {journal} {\bibinfo  {journal} {J. Lond. Math. Soc.}\ }\textbf
  {\bibinfo {volume} {15}},\ \bibinfo {pages} {47} (\bibinfo {year}
  {2007})}\BibitemShut {NoStop}%
\bibitem [{\citenamefont {Thorpe}\ \emph {et~al.}(1973)\citenamefont {Thorpe},
  \citenamefont {Weaire},\ and\ \citenamefont {Alben}}]{Thorpe73}%
  \BibitemOpen
  \bibfield  {author} {\bibinfo {author} {\bibfnamefont {M.~F.}\ \bibnamefont
  {Thorpe}}, \bibinfo {author} {\bibfnamefont {D.}~\bibnamefont {Weaire}},\
  and\ \bibinfo {author} {\bibfnamefont {R.}~\bibnamefont {Alben}},\ }\bibfield
   {title} {\bibinfo {title} {{Electronic Properties of an Amorphous Solid.
  III. The cohesive Energy and the Density of States}},\ }\href
  {https://doi.org/10.1103/PhysRevB.7.3777} {\bibfield  {journal} {\bibinfo
  {journal} {Phys. Rev. B}\ }\textbf {\bibinfo {volume} {7}},\ \bibinfo {pages}
  {3777} (\bibinfo {year} {1973})}\BibitemShut {NoStop}%
\bibitem [{\citenamefont {Haydock}\ \emph {et~al.}(1972)\citenamefont
  {Haydock}, \citenamefont {Heine},\ and\ \citenamefont {Kelly}}]{Haydock72}%
  \BibitemOpen
  \bibfield  {author} {\bibinfo {author} {\bibfnamefont {R.}~\bibnamefont
  {Haydock}}, \bibinfo {author} {\bibfnamefont {V.}~\bibnamefont {Heine}},\
  and\ \bibinfo {author} {\bibfnamefont {M.~J.}\ \bibnamefont {Kelly}},\
  }\bibfield  {title} {\bibinfo {title} {{Electronic structure based on the
  local atomic environment for tight-binding bands}},\ }\href
  {https://doi.org/10.1088/0022-3719/5/20/004} {\bibfield  {journal} {\bibinfo
  {journal} {J. Phys. C}\ }\textbf {\bibinfo {volume} {5}},\ \bibinfo {pages}
  {2845} (\bibinfo {year} {1972})}\BibitemShut {NoStop}%
\bibitem [{\citenamefont {Gaspard}\ and\ \citenamefont
  {Cyrot-Lackmann}(1973)}]{Gaspard73}%
  \BibitemOpen
  \bibfield  {author} {\bibinfo {author} {\bibfnamefont {J.~P.}\ \bibnamefont
  {Gaspard}}\ and\ \bibinfo {author} {\bibfnamefont {F.}~\bibnamefont
  {Cyrot-Lackmann}},\ }\bibfield  {title} {\bibinfo {title} {{Density of states
  from moments. Application to the impurity band}},\ }\href
  {https://doi.org/10.1088/0022-3719/6/21/012} {\bibfield  {journal} {\bibinfo
  {journal} {J. Phys. C}\ }\textbf {\bibinfo {volume} {6}},\ \bibinfo {pages}
  {3077} (\bibinfo {year} {1973})}\BibitemShut {NoStop}%
\bibitem [{\citenamefont {Haydock}\ \emph {et~al.}(1975)\citenamefont
  {Haydock}, \citenamefont {Heine},\ and\ \citenamefont {Kelly}}]{Haydock75}%
  \BibitemOpen
  \bibfield  {author} {\bibinfo {author} {\bibfnamefont {R.}~\bibnamefont
  {Haydock}}, \bibinfo {author} {\bibfnamefont {V.}~\bibnamefont {Heine}},\
  and\ \bibinfo {author} {\bibfnamefont {M.~J.}\ \bibnamefont {Kelly}},\
  }\bibfield  {title} {\bibinfo {title} {{Electronic structure based on the
  local atomic environment for tight-binding bands. II}},\ }\href
  {https://doi.org/10.1088/0022-3719/8/16/011} {\bibfield  {journal} {\bibinfo
  {journal} {J. Phys. C}\ }\textbf {\bibinfo {volume} {8}},\ \bibinfo {pages}
  {2591} (\bibinfo {year} {1975})}\BibitemShut {NoStop}%
\end{thebibliography}
%

\end{document}